\documentclass[letterpaper]{article}
\usepackage{aaai25} % DO NOT CHANGE THIS
\usepackage{times} % DO NOT CHANGE THIS
\usepackage{helvet} % DO NOT CHANGE THIS
\usepackage{courier} % DO NOT CHANGE THIS
\usepackage[hyphens]{url} % DO NOT CHANGE THIS
\usepackage{graphicx} % DO NOT CHANGE THIS
\usepackage{graphicx} % DO NOT CHANGE THIS
\usepackage{natbib} % DO NOT CHANGE THIS
\usepackage{caption} % DO NOT CHANGE THIS
\usepackage[utf8]{inputenc}
\usepackage{url}
\usepackage{algorithm}
\usepackage[noend]{algpseudocode}
\usepackage{booktabs}
\usepackage{xspace}
\usepackage{amsmath, amssymb, amsthm}

\newcommand{\set}[1]{\left\{#1\right\}}

\newcommand{\fpr}[1]{\mathopen{}\left(#1\right)}

\newcommand{\abs}[1]{{\left|#1\right|}}

\newcommand{\np}{\textbf{NP}}
\newcommand{\poly}{\textbf{P}}

\newcommand{\given}[1][]{\:#1\vert\:}
\newcommand{\prbssbm}{\textsc{ssbm}\xspace}
\newcommand{\prbssbmfixed}{\textsc{ssbm-fixed}\xspace}
\newcommand{\prbkclique}{\textsc{$k$-clique}\xspace}

\newtheorem{theorem}{Theorem}
\newtheorem{problem}{Problem}

\algrenewcommand\algorithmicrequire{\textbf{Input:}}

\DeclareRobustCommand{\dispfunc}[2]{%
    \ensuremath{%
        \ifthenelse{\equal{#2}{}}%
            {\mathit{#1}}%
            {\mathit{#1}\fpr{#2}}}}

\newcommand{\bigO}[1]{\dispfunc{\mathcal{O}}{#1}}

\newcommand{\algsingle}[1]{\textsl{Single}}
\newcommand{\algshared}[1]{\textsl{Shared}}
\newcommand{\algmultilevel}[1]{\textsl{Multilevel}}
\newcommand{\algmlsingle}[1]{\textsl{ML+Single}}
\newcommand{\algmlshared}[1]{\textsl{ML+Shared}}

\renewcommand{\refname}{References}
\makeatletter
\renewcommand\bibsection{%
  \section*{{\refname}\@mkboth{\refname}{\refname}}%
}%
\makeatother

\begin{document}

\title{From your Block to our Block: How to Find Shared Structure between\\Stochastic Block Models over Multiple Graphs}
\author {
    % Authors
    Iiro Kumpulainen,\!\textsuperscript{\rm 1}
    Sebastian Dalleiger,\!\textsuperscript{\rm 2}
    Jilles Vreeken,\!\textsuperscript{\rm 3}
    Nikolaj Tatti\textsuperscript{\rm 1}
}
\affiliations {
    % Affiliations
    \textsuperscript{\rm 1}HIIT, University of Helsinki\\
    \textsuperscript{\rm 2}KTH Royal Institute of Technology\\
    \textsuperscript{\rm 3}CISPA Helmholtz Center for Information Security\\
    \{iiro.kumpulainen, nikolaj.tatti\}@helsinki.fi, sdall@kth.se, jv@cispa.de
}

\date{}

\maketitle

\begin{abstract}
Stochastic Block Models (SBMs) are a popular approach to modeling single real-world graphs. The key idea of SBMs is to partition the vertices of the graph into blocks with similar edge densities within, as well as between different blocks. However, what if we are given not one but multiple graphs that are unaligned and of different sizes? How can we find out if these graphs share blocks with similar connectivity structures?
In this paper, we propose the shared stochastic block modeling (SSBM) problem, in which we model $n$ graphs using SBMs that share parameters of $s$ blocks. We show that fitting an SSBM is \np-hard, and consider two approaches to fit good models in practice. In the first, we directly maximize the likelihood of the shared model using a Markov chain Monte Carlo algorithm. In the second, we first fit an SBM for each graph and then select which blocks to share. We propose an integer linear program to find the optimal shared blocks and to scale to large numbers of blocks, we propose a fast greedy algorithm. Through extensive empirical evaluation on synthetic and real-world data, we show that our methods work well in practice.
\end{abstract}

\begin{links}
\link{Code}{https://version.helsinki.fi/dacs/SharedSBM}
\end{links}

\section{Introduction}

In many network settings, we are interested in what is similar and what is different between two or more graphs. Most existing methods, however, focus on {quantifying} similarity rather than {qualifying} what are the structures that are specific to or shared between the graphs. Examples of when we need more than just a measure of \emph{how} similar graphs are, but we are specifically interested in characterizing \emph{what} makes the graphs similar, include comparing brain scans of different patients; protein interaction networks of different species; social networks of different kinds; or comparing trade networks measured at different points of time. In some of these examples, the graphs will have aligned vertices, but often the graphs are unaligned or of different sizes. In this paper, we propose a method for finding shared connectivity structures between graphs that work in each of these settings.

We build our method upon the notion of the Stochastic Block Model (SBM), which is a popular approach to modeling single graphs~\citep{peixoto2019bayesian}. The key idea of the SBM is to partition the vertices of a graph into {blocks}, modeling the edge likelihoods of nodes within a block as well as those of nodes in different blocks using Bernoulli variables. By optimizing likelihood, we can so obtain a partition such that the edge densities within each block are similar, and those between every pair of blocks are similar too.

Towards our goal of revealing shared structure, we extend the notion of the SBM to $n$ graphs, modeling each with its own SBM, but, requiring these $n$ models to share $s$ blocks. That is, we model the edges of a shared block using the {same distribution} across all $n$ graphs. This formulation allows the domain expert to flexibly explore what is similar (the $s$ shared blocks) and what is different (the other blocks) between the $n$ graphs. We dub this the Shared Stochastic Block Model or SSBM for short.

We show that fitting the optimal SSBM, as well as the sub-problem where we know the block memberships per graph, are both \np-hard, even for $n = 2$ graphs. Moreover, these problems are inapproximable unless $\poly=\np$. To fit good models in practice, we propose two approaches. In the first, we jointly search for the optimal blocks that should be shared. We show how we can use a Markov chain Monte Carlo (MCMC) approach with simulated annealing to search for the optimal assignment. 

Our second approach for fitting a good model consists of two stages. In the first stage, we search only for the assignment of vertices into blocks. Once the vertex assignment is fixed, we then select which blocks should be shared. We show that this assignment can be done optimally with an integer linear program (ILP), but also a faster greedy approach where we select that shared block that reduces the log-likelihood the least. We consider and experiment with several strategies for the initial assignment of vertices:
fitting a model to each graph individually, or using the assignment given from the first approach as the starting point.

Through an extensive set of experiments, we show that our methods work well in practice. On synthetic data we show that 1) we obtain the best results if we use MCMC with an agglomeration heuristic to find initial SBMs per graph, and then an ILP to find the $s$-SSBM, that 2) we can use BIC to recover the ground truth number of shared blocks, and that 3) the runtime of our approaches scales linearly in the number of edges. Through a case study, we show our method reveals easily interpretable knowledge of the similarities between brain scans of ADHD patients.

\section{Preliminaries}\label{section:preliminaries}

Let $G=(V,E)$ be a graph with $\abs{V} = N$ nodes and $\abs{E} = M$ edges. In the standard stochastic block model, we have $B$ blocks and a partition vector $\mathbf{b}$ whose entries $b_i \in \set{1,2, \ldots ,B}$ assign each node $v_i$ in the graph to one of the blocks. In addition, we have a $B \times B$ matrix of probabilities $\mathbf{\Theta}$ where an entry $\theta_{ab}$ denotes the probability of an edge between a node in block $a$ and another node in block $b$.

Let $\mathbf{A}$ denote the adjacency matrix of graph $G$ with entries $A_{ij} \in \set{0,1}$ indicating edges between nodes $v_i$ and $v_j$ in the graph. The likelihood of the graph according to the SBM is given by the Bernoulli distribution
\begin{equation*}
P(G \given \mathbf{\Theta},\mathbf{b}) = 
\prod_{i,j,i\neq j} \theta_{b_i,b_j}^{A_{ij}}(1-\theta_{b_i,b_j})^{1-A_{ij}} \; ,
\end{equation*}
where we assume that the graph is directed. In the case of an undirected graph, the product is over pairs of indices $i,j$ with $i<j$ instead. 
To simplify the notation, we define $C_{ij}$
as the number of edges between blocks $i$ and $j$. Similarly, we define $F_{ij}$
as the number of missing edges between $i$ and $j$.  The log-likelihood of the model is then
\begin{align*}
\log{P(G \given \mathbf{\Theta},\mathbf{b})}
&= \sum_{i,j \in [B]} C_{ij} \log{\theta_{ij}}+F_{ij}\log{(1-\theta_{ij})} \; .
\end{align*}

\section{Problem Definition}\label{section:theory}

We argue that stochastic block models are also valuable in settings with multiple graphs and can be used to compare connectivity patterns. Shared models may better explain common structures in the data than individual SBMs and require fewer parameters. To that end, we introduce the concept of shared blocks in SBMs. We then define associated optimization problems and show that they are \np-hard.

%\subsection{Problem Definition}

Assume that we have $n$ graphs $G_1, \ldots, G_n$ with an SBM for each graph $G_k$ with the partition vector $\mathbf{b}_k$ and the probability matrix $\mathbf{\Theta}_k$ with entries $\theta^k_{ij}$. We then say that these SBMs have $s$ shared blocks if we can select $s$ blocks from each SBM such that the probabilities of edges between corresponding block pairs are identical in each SBM. 

More formally, SBMs share $s$ blocks if there are $n$ injective mappings $\sigma_1, \ldots, \sigma_n$, where $\sigma_k: [s] \rightarrow [B_k]$ such that $\theta^k_{\sigma_k(i)\sigma_k(j)}$ does not depend on $k$. In other words, the different $\mathbf{\Theta}_k$ matrices must have identical $s \times s$ submatrices corresponding to the shared blocks.
Note that regardless of having shared blocks between the SBMs, we can compute the log-likelihood for each graph as before, and the total log-likelihood for $n$ graphs is
\begin{equation}
\label{equation:log_P_bernoulli_shared}
\begin{split}
&\log{P(G_1, \ldots, G_n \given \set{\mathbf{\Theta}_k,\mathbf{b}_k}_k)} \\
&\qquad\qquad= \sum_{k = 1}^n\log{P(G_k \given \mathbf{\Theta}_k,\mathbf{b}_k)} \; .
\end{split}
\end{equation}

This gives rise to the optimization problem of inferring a maximum-likelihood stochastic block model with shared blocks. We formally define the problem as follows.
\begin{problem}[\prbssbm]
\label{problem:shared_parameter_sbms_fixed_B}
Given $n$ graphs $G_1, \ldots, G_n$, each graph $G_k$ with fixed number of blocks $B_k$, as well as a non-negative integer $s$, find partition vectors $\mathbf{b}_1, \ldots, \mathbf{b}_n$ and probability matrices $\mathbf{\Theta}_k \in \mathbb{R}^{B_k\times B_k}$ with $s$ shared blocks between them such that the log-likelihood $\log P(G_1, \ldots, G_n \given \set{\mathbf{\Theta}_k,\mathbf{b}_k}_{k = 1}^n)$ is maximized.
\end{problem}

One way to approach this problem is to find block assignments for each graph independently and then find which parameters to share between the SBMs such that the log-likelihood remains as high as possible. This leads us to the subproblem of inferring the shared SBM parameters while the assignment of vertices into blocks is fixed.
\begin{problem}[\prbssbmfixed]
\label{problem:shared_parameters_fixed_sbms}

Given $n$ graphs $G_1, \ldots, G_n$ and $n$ 
partition vectors $\mathbf{b}_1, \ldots, \mathbf{b}_n$, as well as a non-negative integer $s$, find probability matrices $\mathbf{\Theta}_k$ with $s$ shared blocks between them such that the log-likelihood $\log P(G_1, \ldots, G_n \given \set{\mathbf{\Theta}_k,\mathbf{b}_k}_{k = 1}^n)$ is maximized.
\end{problem}

Note that the maximum-likelihood probability matrices $\mathbf{\Theta}_k$ for each graph $G_k$ can be easily computed once the block assignments and shared blocks are known.
More formally, let $C^k_{ij}$ and $F^k_{ij}$ denote the numbers of edges and missing edges for a pair of blocks $i, j \in [B_k]$. 
When $i$ and $j$ are not both shared between graphs, the maximum log-likelihood is achieved by
\begin{equation}
\label{equation:Pmatrix}
\theta^k_{ij} = \frac{C^k_{ij}}{C^k_{ij}+F^k_{ij}} \; .
\end{equation}
Moreover, the optimal parameter between the $i$th and $j$th shared block in $k$th graph is

\begin{equation}
\label{equation:Pshared}
\theta^k_{\sigma_k(i)\sigma_k(j)} = \frac{\sum_{\ell=1}^n C_{\sigma_\ell(i)\sigma_\ell(j)}^\ell}{\sum_{\ell=1}^n C^\ell_{\sigma_\ell(i)\sigma_\ell(j)}+F^\ell_{\sigma_\ell(i)\sigma_\ell(j)}} \; .
\end{equation}
Thus, to solve \prbssbm or \prbssbmfixed, it is sufficient to find the assignments of vertices into blocks as well as which blocks to share, and the probability matrices can then be computed according to Equations~\ref{equation:Pmatrix}-\ref{equation:Pshared}.

%\section{Computational Complexity}

Next we show that both \prbssbm and \prbssbmfixed are \np-hard and inapproximable by a reduction from the well-known problem of finding cliques of size $k$.
\begin{theorem}
\prbssbm and
\prbssbmfixed are \np-hard, even for $n = 2$. Moreover, unless $\poly = \np$, there is no polynomial-time algorithm that always produces a solution for \prbssbm or
\prbssbmfixed with a log-likelihood $\mathit{LLH}$ such that $\mathit{LLH} \geq  \alpha \mathit{OPT}$, where $\mathit{OPT}$ is the log-likelihood of the optimal solution and $\alpha \geq 1$ is any constant.
\end{theorem}

The second result states that \prbssbm and \prbssbmfixed are inapproximable. Note that $\alpha \geq 1$ makes sense since the log-likelihood of a Bernoulli SBM is always non-positive.

\begin{proof}
Let us focus first on \prbssbm.
We will use a reduction from the \np-hard \prbkclique problem, which asks whether a given graph has a clique of size $k$. Given a graph $G$ and integer $k$ as an instance of \prbkclique, we construct an instance of \prbssbm with two graphs by setting $G_1 = G$ and $G_2$ as a complete graph with the same number of vertices. In addition, set $s = k$, and set $B_1 = B_2 = \abs{V(G)}$.

We claim that $G$ has a $k$-clique if and only if the optimal log-likelihood equals $0$.
Assume that there is a $k$-clique. Set the partition vectors $\mathbf{b}_1, \mathbf{b}_2$ such that each vertex belongs to a different block.
Set $\mathbf{\Theta}_i$ to be the adjacency matrix of $G_i$. The log-likelihood in this case is $0$ and the clique provides $k$ shared blocks.

Assume that the optimal log-likelihood is 0. This is only possible if each parameter in $\mathbf{\Theta}_1$ and $\mathbf{\Theta}_2$ is 0 or 1.
Let $W$ be the vertices in $G_1$ in blocks shared with $G_2$. Since $k$ blocks are shared, we have $\abs{W} \geq k$. Since $G_2$ is fully connected, Eq.~\ref{equation:Pshared} implies that any parameter between two shared blocks cannot be 0, so it must be 1. Consequently, any two shared blocks are fully connected in $G_1$, that is, $W$ is a clique.
The case for \prbssbmfixed is essentially the same, except now we fix the vectors $\mathbf{b}_1, \mathbf{b}_2$ such that each vertex is in its own block already in the construction.

In summary, solving \prbssbm or \prbssbmfixed also solves \prbkclique, which means that both problems are \np-hard. Furthermore, the optimal objective value is $0$, which means that an approximation algorithm with any multiplicative approximation guarantee would have to find the exact optimal solution thus solving \prbkclique. 
\end{proof}

\section{Algorithms}%\label{section:algorithms}

Our next step is to consider techniques to fit our model. We will first consider a method that solves the main problem \prbssbm with a Markov chain Monte Carlo (MCMC) approach. Next, we consider a two-step approach in which we first find and fix the block assignments and then find the shared blocks. The latter step can be then optimized with an ILP or with a greedy algorithm. We evaluate different group assignment strategies in our experiments.

\subsection{MCMC Algorithm for \prbssbm}%\label{section:mcmc}

Although exact inference of a maximum-likelihood SBM is computationally infeasible, there are efficient algorithms to find an SBM that fits the data well in practice, such as the MCMC approach~\cite{peixoto2019bayesian}. We modify this approach to suit our model.
For each graph $G_k$, we start with a randomly initialized partition vector $\mathbf{b}_k$ and assign the first $s$ blocks in each graph to be shared such that blocks with corresponding indices are mapped to each other, that is, $\sigma_k(i) = i$ for $i \in [s]$. At each step of the algorithm, we propose a new block assignment $\mathbf{b}_k'$ where one vertex is moved to a new block.

Rather than proposing new blocks uniformly randomly, we use a heuristic described by~\citet{peixoto2019bayesian} where moves to neighboring blocks become more likely. 
This move proposal is then accepted with probability
\[
    %P(\mathbf{b}_k \leftarrow \mathbf{b}_k') =
    \min\left(1, \frac{P(\mathbf{b}_k' \given G_1, \ldots, G_n)^\beta P(\mathbf{b}_k \given \mathbf{b}_k')}{P(\mathbf{b}_k \given G_1, \ldots, G_n)^\beta P(\mathbf{b}_k' \given \mathbf{b}_k)}\right),
\]
where $\beta$ is an inverse temperature parameter used for simulated annealing.
%
\iffalse
inverse temperature parameter that we can set to $1$ to converge towards the true equilibrium distribution $P(\mathbf{b}_k \given G_1, \ldots, G_n)$ for sampling or increase $\beta$ toward infinity to maximize the posterior using greedy local maximization or simulated annealing~\cite{kirkpatrick1983optimization}.

The posterior probabilities for $\mathbf{b}_k$ and $\mathbf{b}_k'$ can be computed using the Bayes' theorem
\[
P(\mathbf{b}_k \given G_1, \ldots, G_n) = \frac{P(G_1, \ldots, G_n \given \mathbf{b}_k)P(\mathbf{b}_k)}{P(G_1, \ldots, G_n)} \; .
\]
Note that the intractable constants $P(G_1, \ldots, G_n)$ cancel out, and so do the priors $P(\mathbf{b}_k)$ if we assume that any block assignments are apriori equally likely. Thus,
\fi
%
Computing the acceptance probability boils down to computing the log-likelihood given in Eq.~\ref{equation:log_P_bernoulli_shared} with the maximum likelihood parameters given in Eqs.~\ref{equation:Pmatrix}--\ref{equation:Pshared}.

Instead of computing the full log-likelihood for each move proposal, we can compute the change in log-likelihood more efficiently by keeping track of the counts of edges between blocks along with the counts of vertices in each block. Then each move proposal and update can be performed in time $\bigO{\text{deg}(v) + B_k}$, where $\text{deg}(v)$ represents the degree of the vertex $v$ being moved and $B_k$ is the number of blocks in that graph. Hence, one full iteration over all the nodes in all graphs takes time $\bigO{\sum_{k=1}^n \abs{E_k}+ \abs{V_k}B_k}$, where $\abs{E_k}$ and $\abs{V_k}$ are the numbers of edges and vertices in graph $G_k$. 

The number of graphs $n$ is a small constant in our experiments, so we use a simpler implementation without storing and continuously updating the counts of total edges or vertices for shared blocks. This leads to a running time of $\bigO{n\sum_{k=1}^n \abs{E_k}+\abs{V_k}B_k}$ per iteration over all the nodes. 

\subsection{Integer Linear Program for \prbssbmfixed}
%\label{section:ilp}

Next, we consider solving \prbssbmfixed. The number of blocks in an SBM is typically much smaller than the number of vertices. In our case, when the numbers of blocks $B_k$ are small, we can solve \prbssbmfixed exactly using an integer linear program (ILP).

Solving \prbssbmfixed amounts to finding $s$ injective functions $\sigma_k$ that maximize the log-likelihood. To express this problem as an ILP, we need to introduce additional notation. 
Assume that we are given $\sigma_k$ and fix $i \in [s]$. Then define a vector $r$ of length $n$ with $r_k = \sigma_k(i)$, that is, $r_k$ indicates the $i$th shared block in graph $G_k$. Instead of searching for $\sigma_k$, we will search for such vectors $r$. For this purpose, let us define the space of such vectors as $\mathcal{T} = [B_1] \times [B_2] \times \cdots \times [B_k]$.

Next, we express the log-likelihood as a linear function over certain variables and then impose conditions on those variables so that the program matches solving \prbssbmfixed.
To that end assume that $i$ or $j$ are not shared blocks in $G_k$. Then the associated log-likelihood between these blocks is
\[
    U_{ij}^k = C^k_{ij}\log \theta^k_{ij} +F^k_{ij}\log(1 - \theta^k_{ij}) \; ,
\]
where $\theta^k_{ij}$ is given in Eq.~\ref{equation:Pmatrix}.
On the other hand, let $r, t \in \mathcal{T}$ be two vectors indicating shared blocks. Then the associated log-likelihood between these shared blocks is
\[
    Q_{rt} = \sum_{k = 1}^n C^k_{r_kt_k} \log \theta_{rt} + F^k_{r_kt_k} \log(1 - \theta_{rt}) \; ,
\]
where
\[
    \theta_{rt} = \frac{\sum_{k = 1}^n C^k_{r_kt_k}}{\sum_{k = 1}^n C^k_{r_kt_k} + F^k_{r_kt_k} }\;.
\]

To state
our ILP, we need the following two sets of binary variables. The first set $z_{kij}$, where $k \in [n]$ and $i, j \in [B_k]$, indicates whether blocks $i$ and $j$ are shared in $G_k$.
The second set $w_{rt}$, where $r, t \in \mathcal{T}$,
indicates if $r$ and $t$ correspond to shared blocks.
The log-likelihood is then equal to
\begin{equation}
\label{eq:ilpcost}
\sum_{r, t \in \mathcal{T}} w_{rt} Q_{rt} + \sum_{k=1}^n \sum_{i,j=1}^{B_k} (1-z_{kij}) U^k_{ij} \; .
\end{equation}
Next, we need to impose constraints on $z_{kij}$ and $w_{rt}$.
To that end, we introduce binary variables $x_{ki\ell}$, indicating that for graph $G_k$, the block $i \in [B_k]$ is the shared $\ell$th block. In other words, $x_{ki\ell}$ will correspond to having $\sigma_k(\ell) = i$. To guarantee this correspondence we require that each graph has exactly one $\ell$th shared block by imposing
\begin{align}
\sum_{i=1}^{B_k} x_{ki\ell} &= 1, & k \in [n],\ \ell \in [s] \;  . \label{ilp_shared}
\end{align}
Now, we introduce an auxiliary binary variable $y_{ki}$, indicating that the block $i\in [B_k]$ for graph $G_k$ is shared. We achieve this by requiring that
\begin{align}
        y_{ki} &= \sum_{\ell=1}^{s} x_{ki\ell}, & k \in [n],\ i\in [B_k] \; .  \label{ilp_ys}
\end{align}
Note that since $y_{ki}$ are binary, the corresponding functions $\sigma_k$ are forced to be injective.
By definition, $z_{kij} = 1$ if and only if $y_{ki} = 1$ and $y_{kj} = 1$. We achieve this by requiring that 
\begin{equation}
\begin{aligned}
        z_{kij} &\leq y_{ki},\  z_{kij} \leq y_{kj}, & k \in [n],\ i,j\in [B_k] \; , \\
        z_{kij} &\geq y_{ki}+y_{kj}-1, & k \in [n],\ i,j\in [B_k] \; . 
\end{aligned}
\label{ilp_zs}
\end{equation}
To constrain $w_{rt}$ we introduce binary variables $c_{r\ell}$ that indicate that from each graph $G_k$, the block $r_k$ is mapped to the shared $\ell$th block. We achieve this by requiring
\begin{equation}
\begin{aligned}
        c_{r\ell} &\leq x_{kr_k\ell},\ &r \in \mathcal{T}, \ k =[n],\   \ell \in [s] \; , \\
        c_{r\ell} &\geq 1 + \sum_{k=1}^{n} (x_{kr_k\ell} - 1),\!\!& r \in \mathcal{T}\;  ,\  \ell \in [s] \; . 
\end{aligned}
\label{ilp_cs}
\end{equation}
In addition, we introduce binary variables $d_r$ indicating that $\sigma_k(r_k) = \ell$ for some $\ell \in [s]$. This is done by requiring
\begin{align}
    d_{r} &= \sum_{\ell=1}^{s}c_{r\ell}, & r \in \mathcal{T} \; . \label{ilp_ds} 
\end{align}
Finally, by definition, $w_{rt} = 1$ if and only if $d_{r} = 1$ and $d_{t} = 1$. We achieve this by requiring that 
\begin{equation}
 \begin{aligned}
        w_{rt} &\leq d_{r},\  w_{rt} \leq d_{t},\ w_{rt} \geq d_{r}+d_{t}-1 & r, t\in \mathcal{T} \; . 
 \end{aligned}
\label{ilp_ws}
\end{equation}
Note that Eq.~\ref{eq:ilpcost} is linear since $U^k_{ij}$ and $Q_{rt}$ are constants. Moreover, the constraints in Eqs.~\ref{ilp_shared}--\ref{ilp_ws} are linear. Hence, maximizing Eq.~\ref{eq:ilpcost} subject to Eqs.~\ref{ilp_shared}--\ref{ilp_ws} can be done with an ILP. Finally, the constraints guarantee that we can construct $\sigma_k$ from $c_{r\ell}$ by setting $\sigma_k(\ell) = r_k$ and that the cost in Eq.~\ref{eq:ilpcost} matches to the log-likelihood.

The number of variables and equations are in $\bigO{\abs{\mathcal{T}}^2} = \bigO{\prod B_k^2}$, which grows exponentially with the number of graphs. Moreover, solving an ILP
can be done in $\log(2h)^{\bigO{h}}$ time, where $h$ is the number of variables~\cite{reis2023subspace}. However, the ILP does not depend on the number of nodes or edges in the graphs, and we can solve it efficiently in practice when the number of graphs is small. In addition, we show an alternative ILP in the Appendix, which does not have an exponential number of variables but does depend on the number of nodes in the graphs. 
For our experiments, the above version is adequate.

\subsection{Greedy Algorithm for \prbssbmfixed}

As an alternative to integer linear programming approaches, we propose a greedy algorithm that iteratively picks shared blocks consisting of one block from each graph. In each iteration, the greedy algorithm chooses the shared block that yields the highest log-likelihood with the previously chosen shared blocks. The pseudocode for the greedy algorithm is given in Algorithm~\ref{alg:greedy}.

\begin{algorithm}
\caption{Greedy algorithm for \prbssbmfixed.}
\label{alg:greedy}
\begin{algorithmic}[1]
\Require number of shared blocks $s$, graphs $G_1, \ldots, G_n$ and their block assignments.
\State $S \gets \emptyset;\ 
\mathcal{T} \gets [B_1] \times \cdots \times [B_n]$
\While{$\abs{S} < s$}
    \State $r \gets $
    \begin{minipage}[t]{6.6cm}
     vector in $\mathcal{T}$ with the smallest decrease in log-likelihood when added to shared blocks $S$   
    \end{minipage}\vspace{1mm}

    \State Add $r$ to $S$
    \State Delete any vector from $\mathcal{T}$ sharing blocks with $r$
\EndWhile
\State \Return $S$
\end{algorithmic}
\end{algorithm}

To compute the change in log-likelihood for a single candidate shared block, we update the parameters for edge probabilities between it and the other currently shared blocks. Calculating the log-likelihood change for each of the $n$ graphs for the at most $\bigO{s}$ updated parameters thus takes time $\bigO{ns}$. In each of the $s$ iterations, we consider $\abs{\mathcal{T}}$ candidates, which leads to a total running time of $\bigO{ns^2B_1 \cdots B_n}$. Note that this is polynomial when the number of graphs $n$ is constant, and does not depend on the number of edges or vertices. However, the number of possible shared blocks grows exponentially with the number of graphs, so future work may design different algorithms for cases where $n$ is large. 
\section{Related Work}\label{section:relatedwork}

The Stochastic Block Model (SBM) has been extensively studied and adapted to capture various complexities in network structures. 
The classic SBM, introduced by \citet{holland1983stochastic}, provides a foundational framework for modeling community structures within networks through probabilistic block assignments. 
This model has since been expanded to address different types of network characteristics, 
including degree heterogeneity through the Degree-Corrected SBM \cite{karrer2011stochastic}, 
the incorporation of temporal dynamics into SBMs enabling the analysis of evolving networks \cite{matias2017statistical}, and
the ability to model multi-layer networks with SBMs \cite{qiuyi15mlsbm}. 

\textbf{Sharing.}\;
Sharing in the context of the SBM framework has focused on capturing overlapping communities. 
\citet{airoldi2008mixed} developed the mixed-membership stochastic block model (MMSBM), which allows nodes to belong to multiple communities. 
While Consensus SBM (CSBMs) \cite{faskowitz2018weighted} aim to create a single unified SBM from a collection of hundreds of individual SBMs, they cannot distinguish between shared and specific blocks.
This approach is informed by methods developed for community detection in multiplex networks and overlapping community models. 
The work of \citet{yang2012community} on overlapping community detection models how nodes can belong to multiple communities simultaneously, which they efficiently put into practice using Non-Negative Matrix Factorization~\citep{yang2013overlapping}. 
Recent work by \citet{addad2024multiview} introduces multi-view SBMs to estimate a shared SBM jointly for multiple graphs. 
To date, research on fine-grained block-sharing in the context of multiple graphs is lacking.

\textbf{Inferring.}\;
Markov Chain Monte Carlo (MCMC) methods are widely used to estimate posterior distributions over block assignments and model parameters~\citep{peixoto2014efficient}. 
\citet{latouche2012variational} pioneered the use of variational inference~\citep{blei17vi} to efficiently approximate the posterior distributions in large-scale SBM applications~\citep{tabouy2020variational}, enhancing computational feasibility and offering a more scalable inference.
\citet{peixoto2017mcsbm} further refined these techniques by employing nonparametric Bayesian inference to better capture complex community structures without requiring predefined numbers of blocks. 
Additionally, clustering techniques, such as spectral clustering~\citep{lei2015consistency}, and graph partition heuristics, such as Kernighan-Lin~\citep{KL1970}, have been used to efficiently estimate block structure.
\citet{funke2019stochastic} provide a comparative study of the experimental efficacy of many inference algorithms. 

\textbf{Comparing Graphs.}\;
Graph kernels~\citep{vishwanathan2010graph, kriege2020survey}, such as the graphlet kernel~\citep{shervashidze09a}, the Weisfeiler-Lehman (WL) kernel~\citep{shervashidze2011weisfeiler}, and Wasserstein WL kernel~\citep{togninalli2019wasserstein}, can be used to compare structural properties of \emph{pre-existing} communities and blocks, but fail to capture the high-level commonalities between distributions and blocks that are specific to our use case.
While divergences and metrics over distributions, such as Wasserstein kernels~\citep{de2020wasserstein}, may perform better for many needs, they also require pre-existing communities which we aim to find. %and a stochastic block model or probabilistic model, and combined with clustering methods like $k$-means to group blocks, they do not come with guarantees.

\section{Experiments}\label{section:experiments}
Next, we present our experimental evaluation.

\textbf{Methodology.}
To discover block assignments---while ignoring the shared block requirement---we consider three approaches: 
\algsingle{} independently fits a single SBM to graph using MCMC. The algorithm samples a new assignment by moving a random node between blocks~\cite{peixoto2019bayesian}. Alternatively,
\algmultilevel{} fits single SBMs for each graph using an MCMC algorithm with an agglomerative heuristic that starts with all nodes in separate blocks and uses a series of block merges and individual node movements to reach a stable node assignment for the desired number of blocks~\cite{peixoto2014efficient}. Third, \algmlsingle{} first runs the \algmultilevel{} algorithm as a starting point for \algsingle{} to refine the node assignment further.

To find block assignments with shared blocks, we consider two approaches:
\algshared{} fits SBMs with shared parameters using the MCMC algorithm for \prbssbm.
Finally, \algmlshared{} uses \algmultilevel{} similarly as a starting point for \algshared{}. However, since \algshared{} assumes that the first $s$ blocks in each graph share parameters, we use the ILP, given in Eqs.~\ref{eq:ilpcost}--\ref{ilp_ws}, to select which blocks to share and reorder them to be first. 
We use the last 3 methods directly to find SBMs with shared blocks. In addition, we use all 5 methods to compare the ILP and Greedy algorithms for \prbssbmfixed, with a \textsl{Random} baseline that randomly selects the shared blocks.

We implemented \algsingle{} and \algshared{} in C++ using graph-tool,\!\footnote{\url{https://graph-tool.skewed.de/}} which includes \algmultilevel{}. Our experiments and other algorithms are implemented in Python, and we use Gurobi for solving the ILP instances.\!\footnote{\url{https://www.gurobi.com/solutions/gurobi-optimizer/}} The experiments were run on an AMD EPYC 7452 32-core processor in a high-performance computing environment. All code for reproducing the experiments is publicly available.\!\footnote{\url{https://version.helsinki.fi/dacs/SharedSBM/}}

\subsection{Experiments on Synthetic Graphs}

For each experiment on synthetic data, we create $n$ graphs using stochastic block models with 200 to 500 vertices with a fixed number of blocks and shared blocks. The vertices are randomly partitioned into the blocks, edge probabilities between blocks are sampled from a Beta distribution with parameters $\alpha=0.5, \beta=1.0$, and the probabilities for shared blocks are set to be equal. Each pair of vertices then has an edge between them based on the edge probability between the corresponding blocks. 

We evaluate how accurately our algorithms for \prbssbmfixed can find the ground truth shared blocks by computing the Shared ARI, which we define as the adjusted Rand index (ARI)~\cite{hubert1985comparing} score between an inferred binary partition of vertices into shared or non-shared blocks and the corresponding ground truth partition. 

\begin{figure}[h!]
    \centering
    \includegraphics[width=\columnwidth]{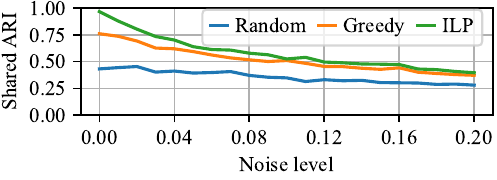}
    \caption{Comparison of ARI scores for the partitions of vertices into shared or non-shared blocks by different algorithms for selecting shared blocks, with increasing levels of random noise in the input block assignment.}
    \label{fig:Shared_ARI_vs_noise_level}
\end{figure}

\textbf{Random Noise.}\; 
Since solving \prbssbm via \prbssbmfixed requires a separate inference algorithm to provide block assignments that may be imperfect, we assess how mistakes in the input block assignment reflect in the Shared ARI scores for our algorithms for optimizing the shared blocks. We add noise to the ground truth block assignment by independently moving vertices to random blocks with increasing probability. Figure~\ref{fig:Shared_ARI_vs_noise_level} shows that ILP outperforms Greedy and Random in finding the ground truth shared blocks when provided with the true assignment of blocks and how increasing levels of random noise slowly degrades performance.

\begin{figure}[h!]
    \centering
    \includegraphics[width=\columnwidth]{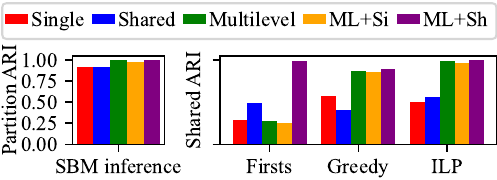}
    \caption{(Left): Average ARI scores measuring the similarity between the inferred partitions of vertices into blocks with the ground truth partitions. (Right): ARI scores between ground truth and inferred partitions of vertices into shared or non-shared blocks for different algorithms for inferring the block assignments paired with different methods for choosing the shared blocks. The Firsts method represents choosing the first $s$ blocks in each graph to be shared, which is effectively random for \algsingle{}, \algmultilevel{}, and \algmlsingle{}.}
    \label{fig:Partition_and_Shared_ARI}
\end{figure}

\textbf{Block Assignment Accuracy.}\; 
We then compare the algorithms for inferring the block assignments by using the ARI score to measure the similarity between the inferred and ground truth partitions of vertices into blocks. Figure~\ref{fig:Partition_and_Shared_ARI} (left) shows the average partition ARI over 10 runs on 3 graphs with 500 vertices and 5 blocks each, with 3 shared blocks between the graphs. All algorithms find partitions highly similar to the ground truth, and we note that \algmultilevel{} is more consistent at finding partitions close to the ground truth than the simpler \algsingle{} and \algshared{} algorithms. \algmlshared{} performs the best by slightly improving upon \algmultilevel{} and finding the ground truth partition nearly every time.

\textbf{Accuracy of Discovering Shared Blocks.}\; 
Combining the inference algorithms with the algorithms for selecting which blocks to share, we obtain a pipeline for solving \prbssbm. Figure~\ref{fig:Partition_and_Shared_ARI} (right) shows the Shared ARI scores for different pairs of algorithms for inferring the block assignments and for choosing the shared blocks. When selecting the $s$ first blocks to be shared, \algshared{} is notably better than random, but the multi-step approaches involving \algmultilevel{} perform better, with results close to ground truth when using ILP. Notably, the \algmlshared{} combination slightly improves upon \algmultilevel{}, while \algmlsingle{} does not.

\begin{figure}[h!]
    \centering
    \includegraphics[width=\columnwidth]{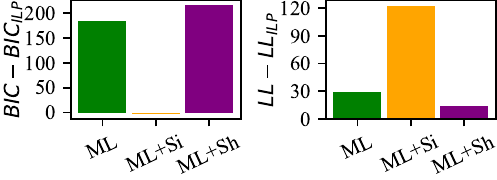}
    \caption{Decrease in BIC (higher is better) and in log-likelihood (lower is better) for \algmultilevel{} (ML), \algmlsingle{} (ML+Si), and \algmlshared{} (ML+Sh) when using ILP shared blocks compared to not using any shared blocks.}
    \label{fig:Decrease_in_BIC_and_Log_Likelihood}
\end{figure}

\textbf{Information Gain.}\;
To evaluate if sharing the parameters between the blocks leads to a better model, we use the Bayesian information criterion (BIC) for model selection. In Figure~\ref{fig:Decrease_in_BIC_and_Log_Likelihood}, we compare the differences in BIC and log-likelihood for \algmultilevel{}, \algmlsingle{}, and \algmlshared{} when using no shared blocks compared to when using the ILP for selecting shared blocks and sharing parameters. When sharing blocks, the BIC slightly decreases for all three, implying that having shared blocks leads to a better model despite small decreases in the log-likelihood. The improvement in BIC is highest for \algmlshared{}, whereas for \algmlsingle{} the loss in log-likelihood nearly offsets the information gain from decreasing the number of parameters.

\begin{figure}[h!]
    \centering
    \includegraphics[width=\columnwidth]{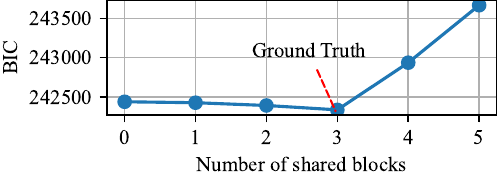}
    \caption{BIC scores for different numbers of shared blocks for \algmlshared{} with ILP. The BIC is lowest at 3 shared blocks matching the ground truth value used for generating the graphs.}
    \label{fig:n_shared_blocks_vs_BIC}
\end{figure}

\textbf{Inferring the Number of Shared Blocks.}\;
While our experiments use fixed numbers of blocks and shared blocks, in practical applications these may have to be inferred from the data. Figure~\ref{fig:n_shared_blocks_vs_BIC} shows the BIC scores for \algmlshared{} with ILP assuming different numbers of shared blocks for two graphs with 400 vertices and 8 blocks each and 3 shared blocks between them. The BIC score is lowest when the assumed number of shared blocks matches the ground truth, showing we can infer it from the data. In future work, the numbers of blocks and shared blocks could be inferred automatically using Bayesian inference~\citep{peixoto2019bayesian}.

\begin{figure}[h!]
    \centering
    \includegraphics[width=\columnwidth]{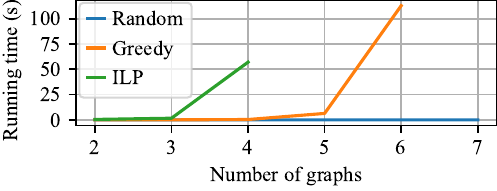}
    \caption{Average running times of different algorithms for optimizing which blocks to share as a function of the number of graphs.}
    \label{fig:shared_block_alg_running_times}
\end{figure}

\textbf{Running Time.}\;
We compare the running times of ILP, Greedy, and Random for an increasing number of graphs, with each graph having 4 blocks, and the number of shared blocks set to 2. We show the results in Fig.~\ref{fig:shared_block_alg_running_times}. Greedy is significantly faster than ILP, taking less than two minutes for six graphs. The exponential trend matches the theory.

\begin{figure}[h!]
    \centering
    \includegraphics[width=\columnwidth]{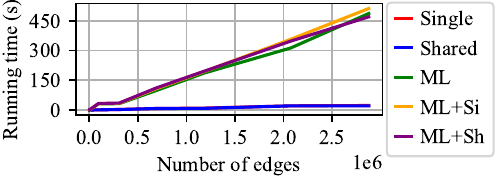}
    \caption{Average running times of different SBM inference algorithms as a function of the number of edges in randomly generated synthetic graphs.}
    \label{fig:inference_running_times_edges}
\end{figure}

Finally, we analyze the total running times for the different MCMC-based inference algorithms. We run \algsingle{} and \algshared{} for 500 iterations over all vertices, while \algmultilevel{} uses graph-tool's default parameters. Figure~\ref{fig:inference_running_times_edges} shows the running times as a function of the total number of edges in two graphs with 4 blocks each and 2 shared blocks. The running times of all algorithms scale linearly with the edge count, but \algsingle{} and \algshared{} are significantly faster, and combining them with \algmultilevel{} adds almost no overhead.

\subsection{Use Case: Brain Networks}
To assess if our algorithms can find shared block structures that can be meaningful in practice, we test them on brain network samples from the NITRC ADHD-200 \cite{BELLEC2017275} resting-state dataset available in the nilearn library.\!\footnote{https://nilearn.github.io/} 
%\!\footnote{https://nilearn.github.io/stable/modules/description/adhd.html} 
We obtain one graph per patient, with each graph having 1000 nodes representing brain regions of interest, and we construct an undirected edge whenever the functional connectivity between the corresponding brain regions has a correlation coefficient of at least 0.5.

\begin{figure}[t!]
    \centering
    \includegraphics[width=\columnwidth]{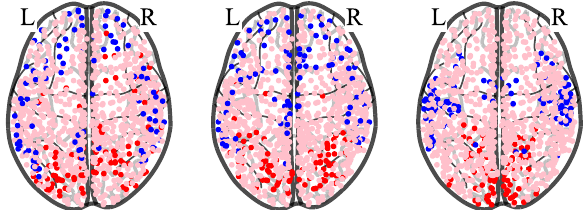}
    \caption{Brain networks for three patients with ADHD at the Oregon Health \& Science University (ages 9, 9, and 11). Brain regions highlighted in red and blue correspond to shared blocks between the graphs having similar connectivity patterns identified by \algmlshared{} with Greedy.}
    \label{fig:adhd_brains}
\end{figure}

We note that the brain networks between patients vary greatly. In Figure~\ref{fig:adhd_brains} we highlight a set of three patients with attention deficit and hyperactivity disorder (ADHD) whose brains display similar connectivity patterns. We set the number of blocks to 10 per graph of which we seek to share 2. 
By using the Greedy algorithm, \algmlshared{} identifies shared blocks that correspond to approximately the same brain region.
Being able to discover shared structures solely based on the connectivity patterns, our method does not use the dataset-inherent node-alignments.

\subsection{Use Case: Wikipedia Networks}
To test the scalability of our methods on real-world networks, we run an experiment on the extensive English and Chinese Wikipedia link networks from the Koblenz network collection~\citep{kunegis2013konect}. Each node represents an article on Wikipedia with directed edges representing links from one article to another. The graph for English Wikipedia has 13~593~032 nodes and 437~217~424 edges, while the Chinese Wikipedia has 1~786 381~nodes and 72~614~837 edges. Since many Wikipedia articles exist in both languages, we expect that the two networks include similar connectivity patterns that shared blocks could represent.

For the large Wikipedia networks, \algmultilevel{} took too long to execute, exceeding our time limit of 8 hours. However, with 100 iterations over all nodes, \algsingle{} and \algshared{} were completed in approximately 5.5 hours. With the number of blocks set at 20 per graph and having 2 shared blocks between the graphs, the ILP took prohibitively long while the Greedy algorithm took less than one second. \algshared{} resulted in a better log-likelihood than \algsingle{} ($-5.036\times 10^9$ versus $-5.050\times 10^9$), and the BIC scores improved in both cases when incorporating the shared blocks chosen by Greedy compared to when not sharing any parameters between blocks. This suggests our models with shared blocks may fit the data better. In this instance, further analysis of the shared blocks would be difficult due to the lack of node labels in the dataset.

\section{Conclusions}

In this paper, we studied the problem of common structure in multiple graphs using the stochastic block model. We showed that the related optimization problems are \np-hard and inapproximable. To fit the model we considered several approaches, based on a Markov chain Monte Carlo (MCMC) approach, integer linear programming, and greedy optimization. In our experiments, the best approach was to first discover block structure using MCMC, and then refine the result by optimizing the shared blocks with an integer linear program. We showed that our algorithms can find shared structure between graphs that may be of practical interest and that 
our methods can be applied to large graphs with over $10^8$ edges.

Developing even better or faster algorithms for finding shared blocks, especially when the number of input graphs is large, is an interesting direction for future work. 
Alternatively, other definitions for shared blocks may be considered, such as allowing the shared blocks to repeat within the graphs by using many-to-many instead of injective mappings to indicate which blocks are shared.
Furthermore, the models for shared blocks could be used in Degree-Corrected SBMs---for example, by requiring that the distributions of node degrees are also shared---or other variants of the SBM.

\section*{Acknowledgements}
This research is supported by the Academy of Finland project MALSOME (343045).
This research is supported by the Wallenberg AI, Autonomous Systems and Software Program (WASP) funded by the Knut and Alice Wallenberg Foundation.

\bibliography{references}

\newpage
\appendix
\section{Appendix --- Alternative Integer Linear program for \prbssbmfixed}

In this section we will describe alternative definition for ILP. This definition avoids having exponential number of parameters but it is most likely impractical due to a large number of parameters required.

The key observation is that the optimal parameters
%given by Eqs.~\ref{equation:Pmatrix}--\ref{equation:Pshared}
are rational numbers of certain forms.
Let $\Lambda$ be the set of all rational numbers $x/y$, where $x, y$ are integers such that 
\[
    0 \leq x \leq y \leq \sum_{k = 1}^n {\abs{V(G_k)} \choose 2}.
\]
We will index $\Lambda$ with $a$, that is, we will write $\Lambda = \set{\lambda_a}$. It follows that $\Lambda$ is of polynomial size and the optimal parameters are in $\Lambda$.

In our ILP we will map each block in each graph to an index.
Let $x_{kir}$ be a binary variable indicating that the $i$th block in $G_k$ is mapped to index $r$. We will use the first $s$ indices as shared, while the remaining indices can be unique. Consequently, we need $q = s + \sum (B_k- s)$ indices, at most.

We force that that each block is mapped to one index with
\begin{align}
\label{eq:altx}
\sum_{r=1}^q x_{kir} & = 1, &k \in [n], i \in [B_k].
\end{align}
We also force that each graph has $s$ shared blocks with
\begin{align}
\sum_{i=1}^{B_k} \sum_{r = 1}^{s} x_{kir} & = s,
& k \in [n]
\end{align}

Next we introduce a binary variable $z_{r\ell a}$ which indicates that block pair with indices $(r, \ell)$ uses $\lambda_a$. Every block pair should use one $\lambda_a$ which we force by
\begin{align}
\sum_a z_{r\ell a} & = 1, & r, \ell \in [q].
\end{align}

Finally, we introduce a binary variable $y_{kija}$ that states that block pair $(i, j)$ in graph $G_k$ uses $\lambda_a$. To make $y$ consistent with $x$ and $z$, we require that
\begin{align}
\label{eq:alty}
y_{kija} & \geq \sum_{r, \ell = 1}^q x_{kir} + x_{kj\ell} + z_{r\ell a} - 2 %, k \in [n], i, j \in [B_k], r, \ell \in [q],
\end{align}
that is $y_{kija} = 1$ if block $i$ is mapped to index $r$ and block $j$ is mapped to index $\ell$ and this index pair using $\lambda_a$.

Finally, we maximize
\begin{equation}
\label{eq:altcost}
    \sum_{k=1}^n \sum_{i,j=1}^{B_k} \sum_a y_{kija}(C^k_{ij}\log \lambda_a +F^k_{ij}\log(1 - \lambda_a)).
\end{equation}

We should point out that we did not need to specify upper bounds for $y_{kija}$ (in addition to the lower bound given in Eq.~\ref{eq:alty}) since the logarithm terms in Eq.~\ref{eq:altcost} are negative.

The constraints and the objective in Eqs.~\ref{eq:altx}--\ref{eq:altcost} are linear so the problem is solvable with ILP. The number of needed variables and equations is in
\[
    \bigO{(\sum_{k = 1}^n B_k)^2 \sum_{k = 1}^n \abs{V(G_k)}^2}.
\]

\end{document}